\documentclass[letterpaper, 10 pt, conference]{ieeeconf}  

\IEEEoverridecommandlockouts                              

\usepackage{cite}
\usepackage{amsmath,amssymb,amsfonts}
\usepackage{algorithmic}
\usepackage{textcomp}
\usepackage{multirow}
\usepackage{flushend}
\usepackage{graphicx}
\usepackage{colortbl}
\usepackage{caption}
\usepackage{capt-of}
\usepackage{balance}
\usepackage{subfigure}
\usepackage{fixltx2e}
\usepackage{setspace}
\usepackage{hyperref}
\usepackage{array}

\title{\LARGE \bf
EEG Cortical Source Feature based Hand Kinematics Decoding using Residual CNN-LSTM Neural Network
}

\author{Anant Jain$^{1}$ and Lalan Kumar$^{2}$
\thanks{$^{1}$Anant Jain is with the Department of Electrical Engineering, Indian Institute of Technology Delhi, New Delhi - 110016, India
        {\tt\small anantjain@ee.iitd.ac.in}}%
\thanks{$^{2}$Lalan Kumar is with the Department of Electrical Engineering, Bharti School of Telecommunication, and, Yardi School of Artificial Intelligence, Indian Institute of Technology Delhi, New Delhi - 110016, India
        {\tt\small lkumar@ee.iitd.ac.in}}%
}

\begin{document}

\maketitle
\thispagestyle{empty}
\pagestyle{empty}

\begin{abstract}
Motor kinematics decoding (MKD) using brain signal is essential to develop Brain-computer interface (BCI) system for rehabilitation or prosthesis devices. Surface electroencephalogram (EEG) signal has been widely utilized for MKD. However, kinematic decoding from cortical sources is sparsely explored. In this work, the feasibility of hand kinematics decoding using EEG cortical source signals has been explored for grasp and lift task. In particular, pre-movement EEG segment is utilized. A residual convolutional neural network (CNN) - long short-term memory (LSTM) based kinematics decoding model is proposed that utilizes motor neural information present in pre-movement brain activity. Various EEG windows at 50 ms prior to movement onset, are utilized for hand kinematics decoding. Correlation value (CV) between actual and predicted hand kinematics is utilized as performance metric for source and sensor domain. The performance of the proposed deep learning model is compared in sensor and source domain. The results demonstrate the viability of hand kinematics decoding using pre-movement EEG cortical source data.
\end{abstract}

\section{INTRODUCTION}

Brain-computer interface (BCI) system utilizes brain activation signals to control external devices. In particular, electroencephalogram (EEG) signals are utilized for non-invasive BCI systems due to mobility and low-cost. EEG-based motor intention decoding is addressed in \cite{aggarwal2022review} for motor classification. Classification based BCI system can also be utilized to command robotic devices \cite{huang2019eeg} and control rehabilitation prosthesis\cite{guo2022ssvep}. However, continuous kinematics decoding based motor intention detection can provide efficient control of BCI systems. EEG-based motor kinematics decoding (MKD) has been employed for upper limb motor tasks\cite{jeong2019trajectory, sosnik2020reconstruction}. The multiple linear regression (mLR) is commonly used for kinematics decoding\cite{sosnik2020reconstruction}. Deep learning based decoding models for MKD are explored in \cite{jeong2020brain, jain2022premovnet}.

EEG signals suffer with the volume conduction effect\cite{edelman2015eeg}, which distorts the neuroelectric signals while transmission from cortical surface to the scalp. EEG source imaging (ESI) is utilized for projecting the scalp EEG potentials onto a cortical source space that represent brain geometry. For BCI, the EEG source localization transforms the input features from "sensor-domain" to "source-domain". In comparison to EEG-based motor intention detection, ESI based motor imagery (MI) classification have achieved better performance\cite{li2019decoding, hou2020novel, giri2021cortical}. However, reliable decoding of motion trajectory from cortical source domain needs to be investigated. ESI based hand motion decoding is addressed for reach-to-target task in \cite{sosnik2021reconstruction} and for snake trajectory tracking in\cite{srisrisawang2022applying}. mLR decoder was utilized for hand, elbow and shoulder trajectories estimation during target-reaching experiment in \cite{sosnik2021reconstruction} with ESI signals. For actual movement execution, an average correlation of $0.36\pm0.13$ was reported across all trajectories. In \cite{srisrisawang2022applying}, hand trajectory estimation was performed using a combination of partial least square (PLS) regression model and square-root unscented Kalman filter (SR-UKF) during random snake trajectory tracking. The highest average correlation of $0.31\pm0.09$ and $0.35\pm0.09$ was reported for hand position and velocity, respectively.

In this work, a residual convolutional neural network - long short-term memory (CNN-LSTM) based kinematics decoding model is proposed for MKD. The source-space based input features are taken for hand kinematics estimation for grasp and lift task. The neural information regarding the motor activity reflects on the motor-cortex region approximately 300ms prior to the movement execution \cite{pancholi2022source}. Hence, various time lag windows are utilized to incorporate the pre-movement motor activation for efficient MKD. The EEG windows are considered at 50 ms prior to the movement onset. Further, the source-domain and sensor-domain MKD results are compared. A comparable performance is observed when deep learning model combined with pre-movement EEG data is utilized. 
\begin{figure*}[ht]
	\centering
	\vspace{0.17 cm}
	\includegraphics[width=0.72\textwidth]{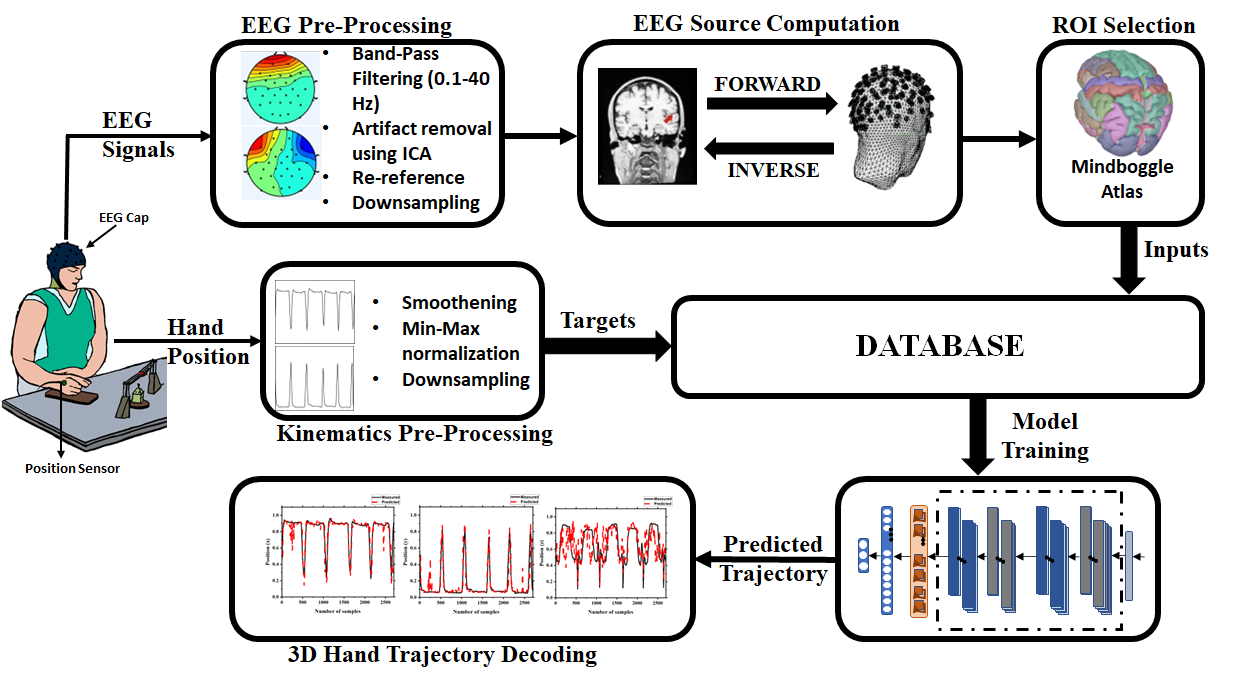}
	\caption{Flowchart of proposed kinematics decoding framework for Grasp-and-Lift task.}
	\label{blockdia} 
\end{figure*}
\section{Methodology}
In this Section, the description of grasp and lift database, data processing steps, EEG cortical source estimation, and deep learning based kinematics estimation framework are included. Fig. \ref{blockdia} shows the flowchart of the proposed kinematics estimation framework.

\subsection{Experimental Setup}

In this work, a publicly available WAY-EEG-GAL\cite{luciw2014multi} database is utilized for hand kinematics estimation. Simultaneous EEG and kinematics data was collected for twelve participants during grasp and lift task. A 32-channel EEG system (Brainproducts ActiCap) is used for EEG signal recording, and 3D hand position is recorded using position sensor. The EEG signals and 3D hand position are recorded with a sampling frequency of 500 Hz. The experiment is performed to grasp and lift the object with different weights and surfaces.

Each trial initiates with turning the LED ON, placed in front of the participant. The participant moves the hand to reach for the object, grasp it, and then lift it for a couple of seconds. With the LED turned OFF, the participant lowers the hand, puts the object on the initial position, and takes the hand back to the resting position.

\subsection{Data Preprocessing}

EEG data is band-pass filtered in the frequency range of 0.1-40 Hz with zero-phase FIR filter to remove baseline drifts. Average re-referencing is employed on the filtered EEG data. Independent component analysis is utilized for removing eye movement artifacts. Data down-sampling form 500 Hz to 100 Hz is performed for the reduction of computational cost. EEGLAB\cite{delorme2004eeglab} toolbox is utilized for EEG data preprocessing.

A zero-phase low-pass FIR filter (cutoff frequency 2 Hz) is utilized for smoothing of hand kinematic data. Further, the smoothed data is normalized in the range of $[0,1]$ using min-max normalization. Normalized data is down-sampled to 100 Hz to match EEG data sampling rate.

\subsection{EEG cortical source Estimation}

Transformation of EEG data from sensor space to source space requires the solution of EEG forward and inverse modeling. Solution of EEG forward modeling gives lead field matrix, that maps the activation at source space with sensor space. The relationship for EEG potential $\mathbf{E}$ for I (=32) sensors, K source dipoles and $N_{s}$ samples, can be expressed as

\begin{equation*} \begin{bmatrix}\mathbf {E}\end{bmatrix}_{I\times N_{s}}=\begin{bmatrix}\mathbf {A}\end{bmatrix}_{I\times K}\,\begin{bmatrix}\mathbf {S}\end{bmatrix}_{K\times N_{s}} + \begin{bmatrix}\mathbf {\eta}\end{bmatrix}_{I\times N_{s}} \tag{1} \end{equation*}
where, $\mathbf{E}$ represents scalp EEG data, $\mathbf{A}$ is lead-field matrix, $\mathbf{S}$ denotes source activity, and $\mathbf{\eta}$ is the measured noise. 

In present study, the numerical Boundary Element Method (BEM) is employed, with ICBM152 MRI template\cite{fonov2011unbiased} as default subject anatomy, to compute head model using OpenMEEG\cite{gramfort2010openmeeg}. EEG inverse modeling solution provides the cortex source distribution based on scalp EEG potential and head model. With the given lead-field matrix, the inverse modeling solution computes $\mathbf{S}$. EEG inverse modeling is a highly under-determined problem, since the number of dipole sources is far more than the scalp EEG sensors. EEG inverse modeling is performed using standardized low-resolution electromagnetic tomography (sLORETA) method\cite{pascual2002standardized}. EEG cortical source estimation for the analysis is performed in Brainstorm toolbox\cite{tadel2011brainstorm}.

\subsection{Mindboggle Atlas}

The brain cortical area is sub-grouped into different regions by incorporating Mindboggle atlas\cite{klein2005mindboggle}. Using Mindboggle atlas, anatomical labels are assigned to cortical structures in MRI data. It labels the cortical brain area into 62 regions. The mean activation of each region is computed to utilize it as input signal for hand kinematics estimation. A total of 62 time-series source signals are obtained for hand kinematics decoding. The selection of regions with Mindboggle atlas is performed using the Brainstorm toolbox\cite{tadel2011brainstorm}. Further, the mean EEG source signals are normalized using z-score normalization.

\subsection{Data Preparation}\label{dp}

For hand kinematics data, the data is segmented from movement onset until the end of the trial. Time lag EEG windows are utilized to consolidate the neural information prior to the actual movement. Various window sizes (150 ms - 300 ms) are utilized for the analysis. EEG cortical source data segment with 150 ms window size will consider EEG data between -200 ms to -50 ms on the time scale, where 0 ms is movement onset. It is to note that pre-movement source data considers 50 ms prior to the actual movement. EEG cortical source matrix with dimension $T\times(M\times N)$ is utilized as input to movement decoder, where $T$ is the data segment, $M$ is total number of scouts, and $N$ is time lag window. For sensor domain analysis, total number of scouts, $M$, is replaced with total EEG sensors ($I$ = 32) in the input matrix.

\subsection{Proposed Residual CNN-LSTM architecture}
\begin{figure}[!t]
	\centering
	\vspace{0.20 cm}
	\subfigure[]{\includegraphics[width=0.24\textwidth]{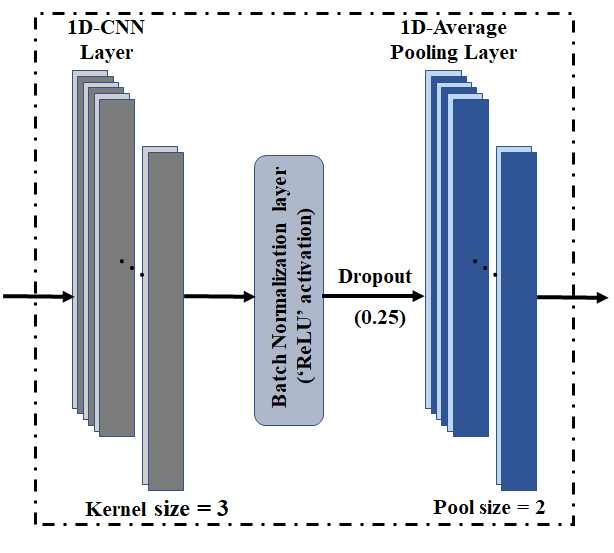}}
	\subfigure[]{\includegraphics[width=0.45\textwidth]{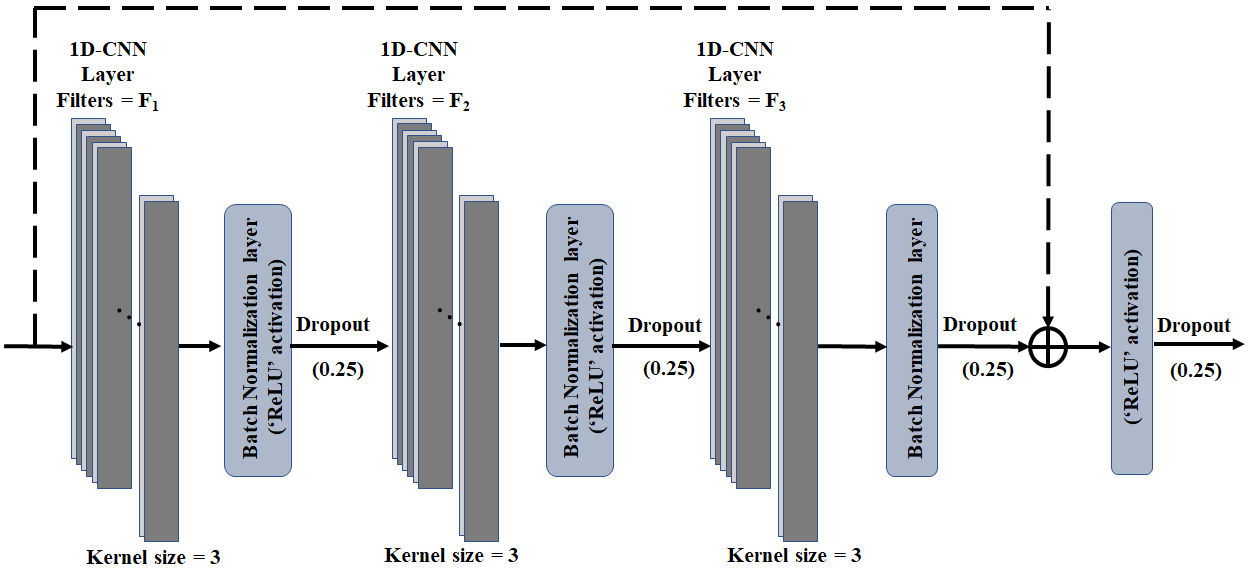}}
	\subfigure[]{\includegraphics[width=0.45\textwidth]{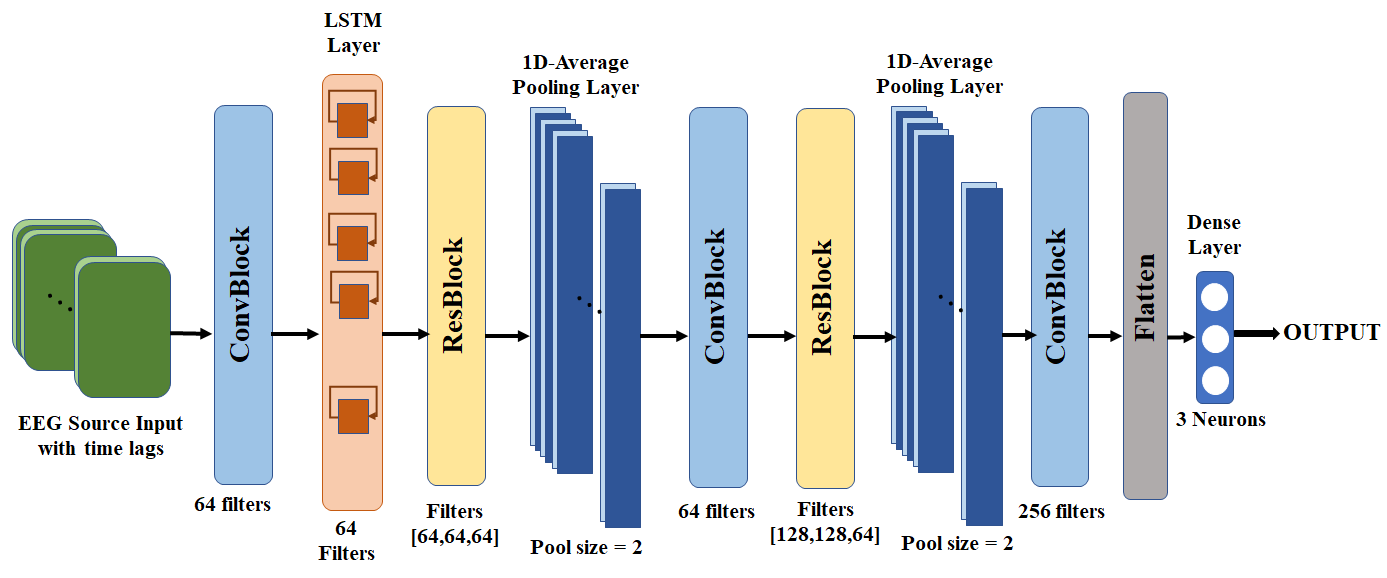}}
	\caption{(a) ConvBlock (b) ResBlock and (c) Proposed residual CNN-LSTM neural decoder}
	\label{resnet}
\end{figure} 
A deep neural structure with residual convolution neural network (CNN) is proposed for hand kinematics estimation. Since, the EEG cortical source signal is time-series signal, 1D-CNN based residual neural network is utilized. The proposed MKD model consists of ConvBlock, LSTM layer and ResBlock. The structures of ConvBlock and ResBlock are shown in Fig. \ref{resnet}(a) and \ref{resnet}(b), respectively. ConvBlock includes a 1D-CNN layer ($C_{1}$ filters and kernel size $=3$), followed by batch normalization layer, ReLU activation layer, drop-out layer (drop rate $=0.5$) and 1D-Average pooling layer (pool size $=3$). ResBlock comprises of three 1D-CNN layers with $[F_{1},F_{2},F_{3}]$ filters, batch normalization layers, drop-out layers and ReLU activation layers, along with a skip connection as shown in Fig. \ref{resnet}(b). The proposed residual CNN-LSTM based neural decoder has been shown in Fig. \ref{resnet}(c). It comprises three ConvBlock, two ResBlock, a LSTM layer, a flatten layer and a dense layer. The dense layer outputs the hand kinematics values in x, y, z-directions. 

\subsection{Training and Evaluation}

For training and performance evaluation of the proposed neural decoders, the database is divided into distinct sets of training, validation and test data. The training and validation data are used to train the model efficiently. The performance evaluation of trained model is performed on test data. The mean square error is used as loss function with adaptive movement estimation (Adam) optimizer for training the decoding model. The early stopping technique (patience$=5$) is utilized to avoid model overfitting. For each participant, a total of 294 trials of grasp-and-lift task are utilized in the analysis from WAY-EEG-GAL dataset. 234, 30 and 30 trials data samples are utilized as training data, validation data, and test data, respectively.

\begin{table}[t]
\centering
\vspace{0.35 cm}
\caption{Hand kinematics decoding analysis of neural decoder with time lag windows of 150 ms, 200 ms, 250 ms, and 300 ms.}
\scalebox{0.51}{
\centering
\begin{tabular}{|c|c||cccc||cccc|}
\hline
\multirow{2}{*}{\textbf{Participants}} & \multirow{2}{*}{\textbf{Directions}} & \multicolumn{4}{c||}{\textbf{Source}}                                                                                                 & \multicolumn{4}{c|}{\textbf{Sensor}}                                                                                                 \\ \cline{3-10} 
                                       &                                      & \multicolumn{1}{c|}{\textbf{150 ms}} & \multicolumn{1}{c|}{\textbf{200 ms}} & \multicolumn{1}{c|}{\textbf{250 ms}} & \textbf{300 ms} & \multicolumn{1}{c|}{\textbf{150 ms}} & \multicolumn{1}{c|}{\textbf{200 ms}} & \multicolumn{1}{c|}{\textbf{250 ms}} & \textbf{300 ms} \\ \hline \hline
\multirow{3}{*}{\textbf{P01}}          & \textbf{x}                           & \multicolumn{1}{c|}{0.55}            & \multicolumn{1}{c|}{0.58}            & \multicolumn{1}{c|}{0.59}            & 0.60            & \multicolumn{1}{c|}{0.68}            & \multicolumn{1}{c|}{0.68}            & \multicolumn{1}{c|}{0.72}            & 0.71            \\ \cline{2-10} 
                                       & \textbf{y}                           & \multicolumn{1}{c|}{0.55}            & \multicolumn{1}{c|}{0.58}            & \multicolumn{1}{c|}{0.59}            & 0.61            & \multicolumn{1}{c|}{0.68}            & \multicolumn{1}{c|}{0.70}            & \multicolumn{1}{c|}{0.73}            & 0.70            \\ \cline{2-10} 
                                       & \textbf{z}                           & \multicolumn{1}{c|}{0.59}            & \multicolumn{1}{c|}{0.50}            & \multicolumn{1}{c|}{0.57}            & 0.63            & \multicolumn{1}{c|}{0.65}            & \multicolumn{1}{c|}{0.66}            & \multicolumn{1}{c|}{0.72}            & 0.66            \\ \hline
\multirow{3}{*}{\textbf{P02}}           & \textbf{x}                           & \multicolumn{1}{c|}{0.46}            & \multicolumn{1}{c|}{0.54}            & \multicolumn{1}{c|}{0.48}            & 0.51            & \multicolumn{1}{c|}{0.61}            & \multicolumn{1}{c|}{0.49}            & \multicolumn{1}{c|}{0.43}            & 0.51            \\ \cline{2-10} 
                                       & \textbf{y}                           & \multicolumn{1}{c|}{0.47}            & \multicolumn{1}{c|}{0.57}            & \multicolumn{1}{c|}{0.52}            & 0.54            & \multicolumn{1}{c|}{0.64}            & \multicolumn{1}{c|}{0.53}            & \multicolumn{1}{c|}{0.47}            & 0.55            \\ \cline{2-10} 
                                       & \textbf{z}                           & \multicolumn{1}{c|}{0.38}            & \multicolumn{1}{c|}{0.54}            & \multicolumn{1}{c|}{0.51}            & 0.52            & \multicolumn{1}{c|}{0.62}            & \multicolumn{1}{c|}{0.55}            & \multicolumn{1}{c|}{0.54}            & 0.59            \\ \hline
\multirow{3}{*}{\textbf{P03}}          & \textbf{x}                           & \multicolumn{1}{c|}{0.46}            & \multicolumn{1}{c|}{0.48}            & \multicolumn{1}{c|}{0.53}            & 0.55            & \multicolumn{1}{c|}{0.45}            & \multicolumn{1}{c|}{0.48}            & \multicolumn{1}{c|}{0.54}            & 0.50            \\ \cline{2-10} 
                                       & \textbf{y}                           & \multicolumn{1}{c|}{0.48}            & \multicolumn{1}{c|}{0.51}            & \multicolumn{1}{c|}{0.56}            & 0.57            & \multicolumn{1}{c|}{0.47}            & \multicolumn{1}{c|}{0.47}            & \multicolumn{1}{c|}{0.56}            & 0.51            \\ \cline{2-10} 
                                       & \textbf{z}                           & \multicolumn{1}{c|}{0.53}            & \multicolumn{1}{c|}{0.56}            & \multicolumn{1}{c|}{0.57}            & 0.57            & \multicolumn{1}{c|}{0.51}            & \multicolumn{1}{c|}{0.49}            & \multicolumn{1}{c|}{0.50}            & 0.46            \\ \hline
\multirow{3}{*}{\textbf{P04}}          & \textbf{x}                           & \multicolumn{1}{c|}{0.70}            & \multicolumn{1}{c|}{0.70}            & \multicolumn{1}{c|}{0.70}            & 0.70            & \multicolumn{1}{c|}{0.73}            & \multicolumn{1}{c|}{0.73}            & \multicolumn{1}{c|}{0.75}            & 0.74            \\ \cline{2-10} 
                                       & \textbf{y}                           & \multicolumn{1}{c|}{0.72}            & \multicolumn{1}{c|}{0.73}            & \multicolumn{1}{c|}{0.73}            & 0.72            & \multicolumn{1}{c|}{0.75}            & \multicolumn{1}{c|}{0.75}            & \multicolumn{1}{c|}{0.76}            & 0.76            \\ \cline{2-10} 
                                       & \textbf{z}                           & \multicolumn{1}{c|}{0.64}            & \multicolumn{1}{c|}{0.63}            & \multicolumn{1}{c|}{0.64}            & 0.63            & \multicolumn{1}{c|}{0.67}            & \multicolumn{1}{c|}{0.65}            & \multicolumn{1}{c|}{0.67}            & 0.66            \\ \hline
\multirow{3}{*}{\textbf{P05}}          & \textbf{x}                           & \multicolumn{1}{c|}{0.47}            & \multicolumn{1}{c|}{0.49}            & \multicolumn{1}{c|}{0.51}            & 0.48            & \multicolumn{1}{c|}{0.53}            & \multicolumn{1}{c|}{0.55}            & \multicolumn{1}{c|}{0.54}            & 0.57            \\ \cline{2-10} 
                                       & \textbf{y}                           & \multicolumn{1}{c|}{0.48}            & \multicolumn{1}{c|}{0.52}            & \multicolumn{1}{c|}{0.55}            & 0.52            & \multicolumn{1}{c|}{0.56}            & \multicolumn{1}{c|}{0.60}            & \multicolumn{1}{c|}{0.60}            & 0.62            \\ \cline{2-10} 
                                       & \textbf{z}                           & \multicolumn{1}{c|}{0.58}            & \multicolumn{1}{c|}{0.61}            & \multicolumn{1}{c|}{0.56}            & 0.56            & \multicolumn{1}{c|}{0.61}            & \multicolumn{1}{c|}{0.60}            & \multicolumn{1}{c|}{0.56}            & 0.57            \\ \hline
\multirow{3}{*}{\textbf{P06}}          & \textbf{x}                           & \multicolumn{1}{c|}{0.37}            & \multicolumn{1}{c|}{0.44}            & \multicolumn{1}{c|}{0.47}            & 0.53            & \multicolumn{1}{c|}{0.49}            & \multicolumn{1}{c|}{0.52}            & \multicolumn{1}{c|}{0.59}            & 0.61            \\ \cline{2-10} 
                                       & \textbf{y}                           & \multicolumn{1}{c|}{0.38}            & \multicolumn{1}{c|}{0.46}            & \multicolumn{1}{c|}{0.48}            & 0.54            & \multicolumn{1}{c|}{0.49}            & \multicolumn{1}{c|}{0.51}            & \multicolumn{1}{c|}{0.57}            & 0.59            \\ \cline{2-10} 
                                       & \textbf{z}                           & \multicolumn{1}{c|}{0.40}            & \multicolumn{1}{c|}{0.45}            & \multicolumn{1}{c|}{0.46}            & 0.49            & \multicolumn{1}{c|}{0.48}            & \multicolumn{1}{c|}{0.47}            & \multicolumn{1}{c|}{0.51}            & 0.53            \\ \hline
\multirow{3}{*}{\textbf{P07}}          & \textbf{x}                           & \multicolumn{1}{c|}{0.41}            & \multicolumn{1}{c|}{0.48}            & \multicolumn{1}{c|}{0.52}            & 0.53            & \multicolumn{1}{c|}{0.55}            & \multicolumn{1}{c|}{0.57}            & \multicolumn{1}{c|}{0.59}            & 0.59            \\ \cline{2-10} 
                                       & \textbf{y}                           & \multicolumn{1}{c|}{0.42}            & \multicolumn{1}{c|}{0.50}            & \multicolumn{1}{c|}{0.52}            & 0.54            & \multicolumn{1}{c|}{0.56}            & \multicolumn{1}{c|}{0.58}            & \multicolumn{1}{c|}{0.61}            & 0.59            \\ \cline{2-10} 
                                       & \textbf{z}                           & \multicolumn{1}{c|}{0.53}            & \multicolumn{1}{c|}{0.53}            & \multicolumn{1}{c|}{0.54}            & 0.52            & \multicolumn{1}{c|}{0.61}            & \multicolumn{1}{c|}{0.61}            & \multicolumn{1}{c|}{0.58}            & 0.54            \\ \hline
\multirow{3}{*}{\textbf{P08}}          & \textbf{x}                           & \multicolumn{1}{c|}{0.58}            & \multicolumn{1}{c|}{0.67}            & \multicolumn{1}{c|}{0.72}            & 0.72            & \multicolumn{1}{c|}{0.68}            & \multicolumn{1}{c|}{0.72}            & \multicolumn{1}{c|}{0.75}            & 0.74            \\ \cline{2-10} 
                                       & \textbf{y}                           & \multicolumn{1}{c|}{0.59}            & \multicolumn{1}{c|}{0.68}            & \multicolumn{1}{c|}{0.73}            & 0.74            & \multicolumn{1}{c|}{0.71}            & \multicolumn{1}{c|}{0.73}            & \multicolumn{1}{c|}{0.76}            & 0.76            \\ \cline{2-10} 
                                       & \textbf{z}                           & \multicolumn{1}{c|}{0.64}            & \multicolumn{1}{c|}{0.63}            & \multicolumn{1}{c|}{0.65}            & 0.65            & \multicolumn{1}{c|}{0.69}            & \multicolumn{1}{c|}{0.69}            & \multicolumn{1}{c|}{0.70}            & 0.70            \\ \hline
\multirow{3}{*}{\textbf{P09}}          & \textbf{x}                           & \multicolumn{1}{c|}{0.56}            & \multicolumn{1}{c|}{0.62}            & \multicolumn{1}{c|}{0.66}            & 0.69            & \multicolumn{1}{c|}{0.64}            & \multicolumn{1}{c|}{0.65}            & \multicolumn{1}{c|}{0.71}            & 0.71            \\ \cline{2-10} 
                                       & \textbf{y}                           & \multicolumn{1}{c|}{0.58}            & \multicolumn{1}{c|}{0.64}            & \multicolumn{1}{c|}{0.69}            & 0.72            & \multicolumn{1}{c|}{0.65}            & \multicolumn{1}{c|}{0.71}            & \multicolumn{1}{c|}{0.74}            & 0.75            \\ \cline{2-10} 
                                       & \textbf{z}                           & \multicolumn{1}{c|}{0.53}            & \multicolumn{1}{c|}{0.55}            & \multicolumn{1}{c|}{0.55}            & 0.54            & \multicolumn{1}{c|}{0.58}            & \multicolumn{1}{c|}{0.54}            & \multicolumn{1}{c|}{0.56}            & 0.59            \\ \hline
\multirow{3}{*}{\textbf{P10}}          & \textbf{x}                           & \multicolumn{1}{c|}{0.67}            & \multicolumn{1}{c|}{0.47}            & \multicolumn{1}{c|}{0.51}            & 0.57            & \multicolumn{1}{c|}{0.53}            & \multicolumn{1}{c|}{0.55}            & \multicolumn{1}{c|}{0.54}            & 0.59            \\ \cline{2-10} 
                                       & \textbf{y}                           & \multicolumn{1}{c|}{0.70}            & \multicolumn{1}{c|}{0.50}            & \multicolumn{1}{c|}{0.54}            & 0.61            & \multicolumn{1}{c|}{0.55}            & \multicolumn{1}{c|}{0.59}            & \multicolumn{1}{c|}{0.61}            & 0.66            \\ \cline{2-10} 
                                       & \textbf{z}                           & \multicolumn{1}{c|}{0.51}            & \multicolumn{1}{c|}{0.46}            & \multicolumn{1}{c|}{0.47}            & 0.48            & \multicolumn{1}{c|}{0.49}            & \multicolumn{1}{c|}{0.51}            & \multicolumn{1}{c|}{0.48}            & 0.50            \\ \hline
\multirow{3}{*}{\textbf{P11}}          & \textbf{x}                           & \multicolumn{1}{c|}{0.54}            & \multicolumn{1}{c|}{0.54}            & \multicolumn{1}{c|}{0.57}            & 0.55            & \multicolumn{1}{c|}{0.59}            & \multicolumn{1}{c|}{0.58}            & \multicolumn{1}{c|}{0.61}            & 0.64            \\ \cline{2-10} 
                                       & \textbf{y}                           & \multicolumn{1}{c|}{0.57}            & \multicolumn{1}{c|}{0.59}            & \multicolumn{1}{c|}{0.61}            & 0.61            & \multicolumn{1}{c|}{0.64}            & \multicolumn{1}{c|}{0.63}            & \multicolumn{1}{c|}{0.66}            & 0.69            \\ \cline{2-10} 
                                       & \textbf{z}                           & \multicolumn{1}{c|}{0.52}            & \multicolumn{1}{c|}{0.50}            & \multicolumn{1}{c|}{0.52}            & 0.53            & \multicolumn{1}{c|}{0.62}            & \multicolumn{1}{c|}{0.61}            & \multicolumn{1}{c|}{0.63}            & 0.65            \\ \hline
\multirow{3}{*}{\textbf{P12}}          & \textbf{x}                           & \multicolumn{1}{c|}{0.50}            & \multicolumn{1}{c|}{0.54}            & \multicolumn{1}{c|}{0.57}            & 0.61            & \multicolumn{1}{c|}{0.55}            & \multicolumn{1}{c|}{0.58}            & \multicolumn{1}{c|}{0.61}            & 0.60            \\ \cline{2-10} 
                                       & \textbf{y}                           & \multicolumn{1}{c|}{0.51}            & \multicolumn{1}{c|}{0.56}            & \multicolumn{1}{c|}{0.61}            & 0.64            & \multicolumn{1}{c|}{0.58}            & \multicolumn{1}{c|}{0.60}            & \multicolumn{1}{c|}{0.64}            & 0.62            \\ \cline{2-10} 
                                       & \textbf{z}                           & \multicolumn{1}{c|}{0.58}            & \multicolumn{1}{c|}{0.61}            & \multicolumn{1}{c|}{0.63}            & 0.64            & \multicolumn{1}{c|}{0.63}            & \multicolumn{1}{c|}{0.66}            & \multicolumn{1}{c|}{0.65}            & 0.65            \\ \hline
\multirow{3}{*}{\textbf{Mean}}         & \textbf{x}                           & \multicolumn{1}{c|}{0.52}            & \multicolumn{1}{c|}{0.55}            & \multicolumn{1}{c|}{0.57}            & \textbf{0.59}            & \multicolumn{1}{c|}{0.59}            & \multicolumn{1}{c|}{0.59}            & \multicolumn{1}{c|}{0.62}            & \textbf{0.62}            \\ \cline{2-10} 
                                       & \textbf{y}                           & \multicolumn{1}{c|}{0.54}            & \multicolumn{1}{c|}{0.57}            & \multicolumn{1}{c|}{0.60}            & \textbf{0.61}            & \multicolumn{1}{c|}{0.61}            & \multicolumn{1}{c|}{0.62}            & \multicolumn{1}{c|}{0.64}            & \textbf{0.65}            \\ \cline{2-10} 
                                       & \textbf{z}                           & \multicolumn{1}{c|}{0.54}            & \multicolumn{1}{c|}{0.55}            & \multicolumn{1}{c|}{0.56}            & \textbf{0.56}            & \multicolumn{1}{c|}{0.59}            & \multicolumn{1}{c|}{0.59}            & \multicolumn{1}{c|}{0.59}            & \textbf{0.59}            \\ \hline
\end{tabular}}
\label{tab:tab01}
\end{table}
\section{Results}
\begin{figure*}[h]
	\centering
	\subfigure[]{\includegraphics[width=0.23\textwidth]{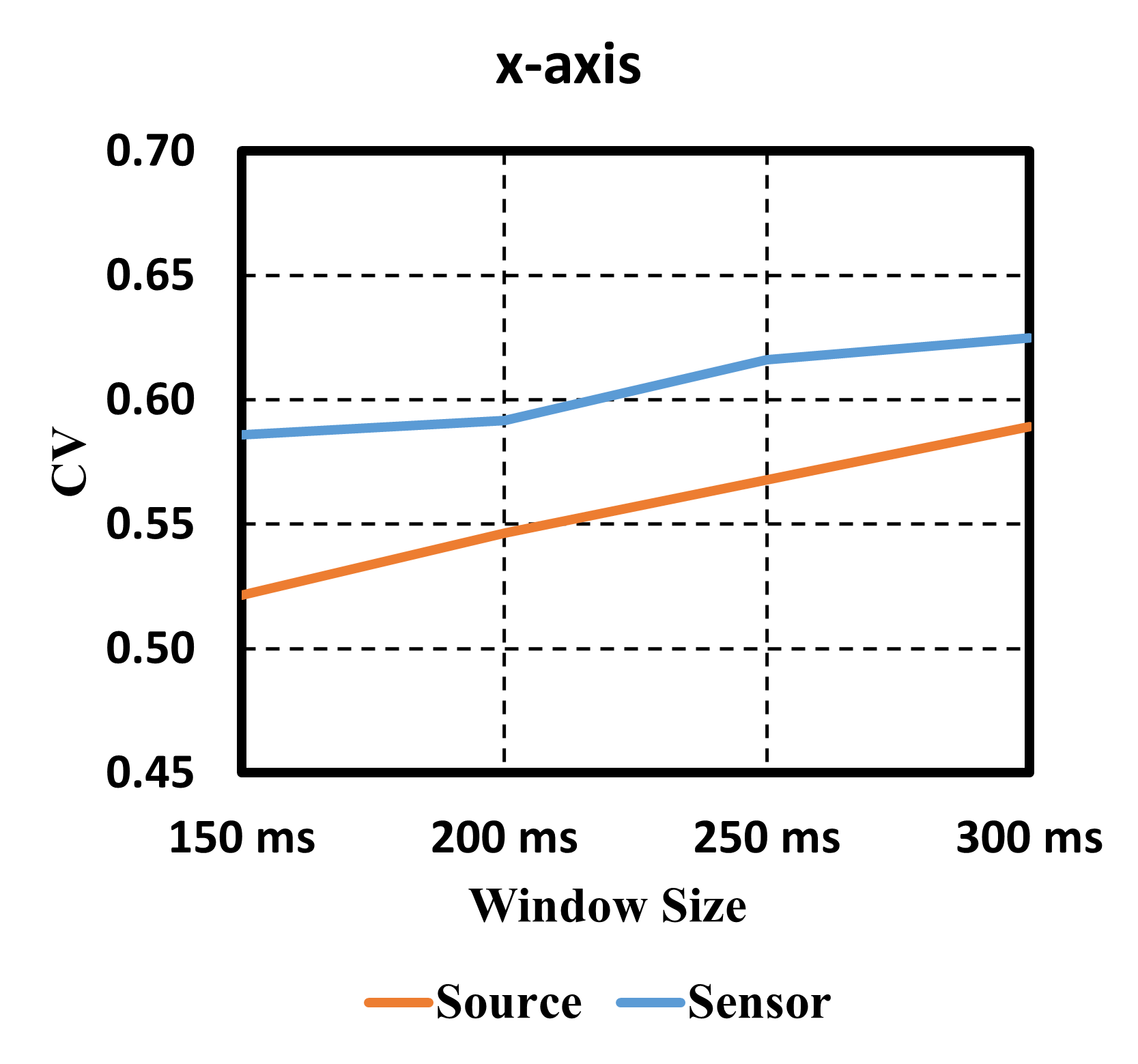}}
	\subfigure[]{\includegraphics[width=0.23\textwidth]{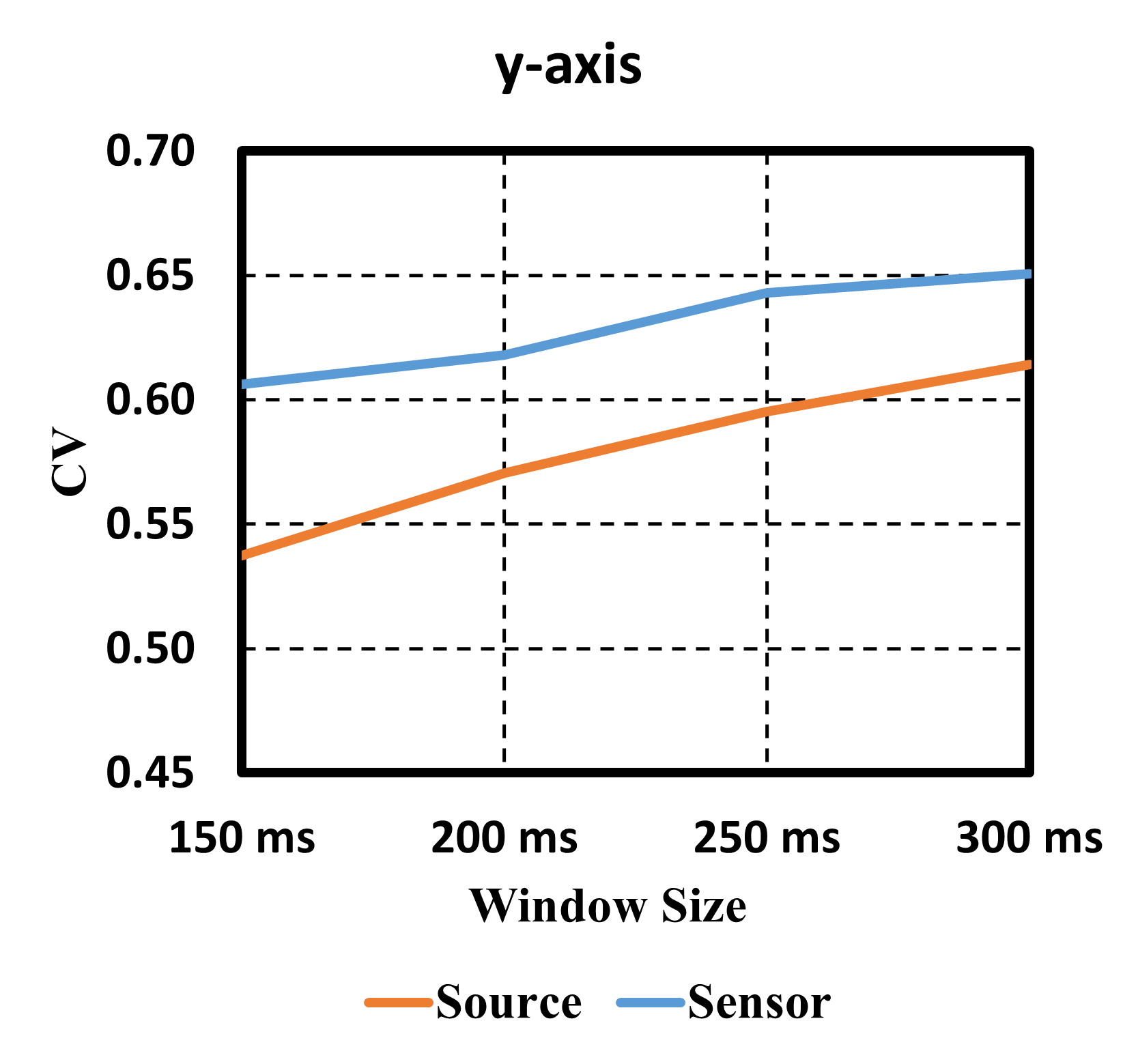}}
	\subfigure[]{\includegraphics[width=0.23\textwidth]{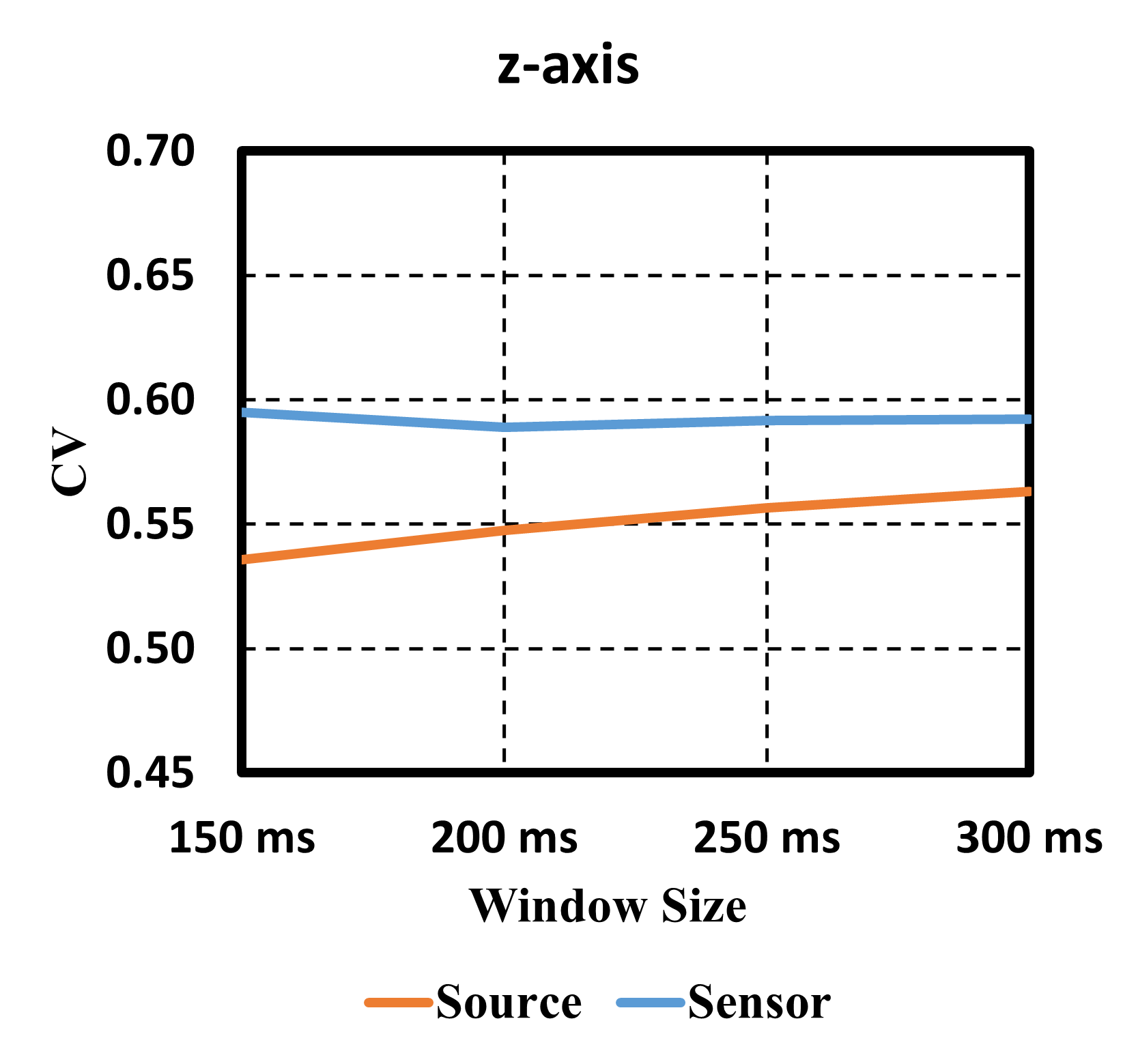}}
	\caption{Hand kinematics decoding performance comparison of source-domain and sensor-domain features for various window sizes in x, y and z-directions.}
	\label{comp}
\end{figure*} 
Correlation value (CV) is taken as performance metric to measure efficiency of the proposed decoding model. The CVs are presented in Table \ref{tab:tab01} for different EEG windows. EEG input to the model is taken as detailed in Section \ref{dp}. Sensor-domain analysis makes use of 32-channel EEG as input. The mean CV is computed for each time lag window. For source-domain analysis, the proposed residual CNN-LSTM decoding model has overall best performance with the window size of 300 ms. The mean CV of the proposed decoding model with 300 ms time lag window are $0.59\pm0.07$, $0.61\pm0.07$, and $0.56\pm0.06$ in x, y, and z-directions, respectively. It is to note that maximum correlation reported for movement execution in source domain is $0.36$ for a target-reaching task\cite{sosnik2021reconstruction}. Application of deep learning model combined with pre-movement EEG data contribute to the performance enhancement in the present work. For sensor-domain analysis, the best decoding performance is found with 300 ms window size with mean CV of $0.62\pm0.08$, $0.65\pm0.08$, and $0.59\pm0.07$ in x, y, and z-directions, respectively. The mean correlation is plotted in Fig. \ref{comp}. It may be noted that MKD performance in source domain is comparable with sensor domain counterparts at larger window size. 
\section{Discussion and Conclusions}

In this study, a residual CNN-LSTM based neural decoder is proposed for kinematics decoding using pre-movement neural information in source domain. WAY-EEG-GAL dataset is utilized for this purpose. The performance evaluation of the proposed model is presented using correlation values (CVs) between actual and predicted trajectories in source and sensor domain. The mean activation of the labeled cortical regions is taken as source-domain features for kinematics decoding. A significant correlation is observed in both the domain with the proposed decoding model. Further improvement needs to be explored for better kinematics decoding. 










\bibliographystyle{IEEEtran}

\bibliography{root}

\end{document}